%% file: main.tex
\documentclass[10pt]{llncs} 

\usepackage{xspace}
\usepackage[latin1]{inputenc}
\usepackage{float}
\newfloat{lstfloat}{pthb}{lop}
\floatname{lstfloat}{Listing}

\usepackage{a4wide} 
\usepackage{footmisc}
\usepackage{bbding}
\usepackage{graphicx}
\usepackage[svgnames, table]{xcolor}
\usepackage{listings}
\usepackage{color, colortbl}
\usepackage{amsmath,amssymb}
\usepackage{booktabs}
\usepackage{xcolor}
\usepackage{tcolorbox}
\usepackage{subcaption}
\usepackage[hyphens]{url} 
\usepackage{wrapfig}

\usepackage{pgf}
\usepackage{tikz}
\usetikzlibrary{arrows,automata}
\usetikzlibrary{positioning}

\input{listings-dafny}

\def\ie{{i.e.},~}
\def\eg{{e.g.},~}

\def\z3{{\sc Z3}\xspace}

\newcommand{\figref}[1]{\figurename~\ref{#1}}

\colorlet{vert}{green!70!black}
\colorlet{rouge}{red!70!black}
\colorlet{orange}{orange!100!black}
\colorlet{bleu}{cyan!80!white!80!black}
\colorlet{gris}{black!10!white}

\newcounter{mynote}
\newlength\mynotewidth
\newcommand{\FC}[1]{%
  \stepcounter{mynote}%
  \raisebox{.36em}{\colorbox{green!20!white}{\textcolor{black}{\tiny\textsf{\S$^{\themynote}$}}}}%
  \marginpar{%
    \sffamily%
    \hspace*{-0.2cm}
    \begin{tabular}{l}
     \colorbox{green!30!white}{\color{green!80!black}\parbox{\mynotewidth}{\small\bfseries{\S~{\themynote}}\hfill
       [Franck]}} \\
    \colorbox{green!50!white}{\color{black}\parbox{\mynotewidth}{\fontsize{7}{8}\selectfont
        #1}}
    \end{tabular}
  }
}

\newcommand{\DJP}[1]{%
  \stepcounter{mynote}%
  \raisebox{.36em}{\colorbox{purple!20!white}{\textcolor{black}{\tiny\textsf{\S$^{\themynote}$}}}}%
  \marginpar{%
    \sffamily%
    \hspace*{-0.2cm}
    \begin{tabular}{l}
     \colorbox{purple!30!white}{\color{purple!80!black}\parbox{\mynotewidth}{\small\bfseries{\S~{\themynote}}\hfill
       [Dave]}} \\
    \colorbox{purple!50!white}{\color{black}\parbox{\mynotewidth}{\fontsize{7}{8}\selectfont
        #1}}
    \end{tabular}
  }
}

\newcommand{\changes}[1]{{\color{red!80!white}{#1}}}
\renewcommand{\changes}[1]{#1}

\makeatletter
\RequirePackage[bookmarks,unicode,colorlinks=true]{hyperref}%
   \def\@citecolor{blue}%
   \def\@urlcolor{blue}%
   \def\@linkcolor{blue}%

\def\orcidID#1{\smash{\href{http://orcid.org/#1}{\protect\raisebox{-1.25pt}{\protect\includegraphics[height=1em]{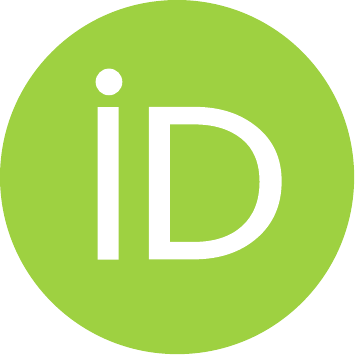}}}}}
\makeatother

\def\dafny{\textsc{Dafny}\xspace}

\AtBeginDocument{}

\begin{document}

\title{\LARGE \bf Formal and Executable Semantics of the Ethereum Virtual Machine in Dafny}
\titlerunning{Formal and Executable Semantics of the EVM in Dafny}

\author{
  Franck Cassez \orcidID{0000-0002-4317-5025}  \and
  Joanne Fuller \and 
  Milad K. Ghale \and
  David J. Pearce\orcidID{0000-0003-4535-9677} \and
  Horacio M. A. Quiles
}

\authorrunning{Cassez, Fuller, Ghale, Pearce, Quiles}

\institute{
    ConsenSys, New York, USA\\
    \email{firstname.surname@consensys.net}
  }

\bibliographystyle{splncs04}

\pagestyle{plain}
\maketitle

\begin{abstract}
The Ethereum protocol implements a replicated state machine.  
The network participants keep track of the system state by: 1) agreeing on the sequence of transactions to be processed and 2) computing the state transitions that correspond to the sequence of transactions.
Ethereum transactions are programs, called \emph{smart contracts}, 
and computing a state transition requires executing some code.
The Ethereum Virtual Machine (EVM) provides this capability and can execute programs written in EVM \emph{bytecode}.
We present a formal and executable semantics of the EVM written in the verification-friendly language \dafny:
it provides $(i)$ a readable, formal and verified specification of the semantics of the EVM; $(ii)$ a framework to formally reason about bytecode. 
\end{abstract}

\input{intro}
\input{background}
\input{formalisation}
\input{evaluation}

\input{relwork}




\section{Conclusion}\label{sec-conclusion}
\input{conclusion}

\end{document}

%% file: listings-dafny.tex
\definecolor{gray}{RGB}{215,215,215}

\lstdefinelanguage{dafny}{
  sensitive=true,
  keywords={},
  otherkeywords={
  <,>, <==>, <=, >=, |, ==, :=},
  basicstyle=\fontsize{10}{11}\selectfont\ttfamily,
  keywordstyle=[2]{\bfseries\color{blue!80!black}},
  commentstyle=\small\rmfamily\itshape \color{orange}, 
  stringstyle=\small\itshape, 
  keywords = [2]{u8, u256, bv256, as, in, module, var, trait, datatype,, type, extends, const, address, constructor, predicate, reads, this, map, modifies, function, method, class, lemma, method, ghost, if, then, else, ensures, requires, decreases, while, do, return, od, assert, assume, expect, invariant, int, nat, bool, Msg, seq, tail, init, last, first, take, returns, datatype},
  moredelim=*[s][commentstyle]{/*}{*/}, 
  morecomment=[l][commentstyle]{//},      
  backgroundcolor=\color{gray!50},
  frame=single, 
  frameround=tttt,
  framesep=0.1cm,
  rulecolor=\color{gray!50},
  moredelim=[is][\ttfamily]{<code>}{</code>}, 
  tabsize=4,
  tab=\rightarrowfill,
  xleftmargin=0.55cm,
  xrightmargin=0.25cm,
  showspaces=false,
  showtabs=false,
  columns=fullflexible,
  numberstyle=\scriptsize,
  keepspaces=true
  morestring=[b]",
  numbers=left,
}

\newcommand{\wlstinline}[1]{\tcbox[tcbox raise
  base,colback=gray!50,colframe=black,nobeforeafter,boxrule=0.25pt,arc=1.5pt,left=0mm,right=0mm,top=0mm,bottom=0mm]{\lstinline[language=dafny,keywords={}]|#1|}}

\renewcommand{\wlstinline}[1]{\lstinline[language=dafny,keywords={}]{#1}}


%% file: intro.tex
\section{Introduction}\label{sec-intro}

A distinctive feature of Ethereum is that transactions are programs, \emph{smart contracts}, and computing a state transition requires to run the contract code to compute the next state.
This capability is provided by the Ethereum Virtual Machine (EVM) that can execute programs written in EVM \emph{bytecode}.
The original and informal specification of the EVM is in the Yellow Paper~\cite{yp-22}.

\smallskip 

As a decentralised platform, Ethereum encourages \emph{client diversity}: network participants are free to choose which implementation of the EVM they want to run, and
there are several implementations to choose from written in different languages \eg Go, Java.
All the EVM implementations must agree on the state transitions, otherwise the network 
would split and the blockchain would \emph{fork}. 
However, the original specification in the Yellow Paper~\cite{yp-22} has some known shortcomings:  
($i$) it is hard to read and does not provide a formal semantics of the EVM and the bytecode;
($ii$) the lack of a formal semantics makes it hard for Ethereum client developers to guarantee that they interpret the Yellow Paper in a consistent way;
($iii$) designing compilers from high-level languages (\eg Solidity\footnote{The most popular language to write smart contracts.}) to EVM bytecode without a formal semantics is error-prone
and, without a precise semantics of the EVM, it is hard to design \emph{certified}
\changes{compilers (preserving of semantics from a high to a low-level language.)}.


One can argue that existing implementations of the EVM (\eg in Go,
Java) provide a {\em de facto} semantics for it.
Whilst this is true to some
extent, such implementations do not enable \emph{formal reasoning}
about bytecode.  Furthermore, whilst smart contracts can be written in
high-level languages like Solidity, they must be compiled into EVM
bytecode before being executed on the EVM.  Tools for checking safety
properties (\eg absence of overflow, division by zero, etc) at the
Solidity level are problematic if they cannot guarantee such
properties hold at the bytecode level.
%
One solution is to design a
provably correct compiler, but this is a complex and long-term
endeavour~\cite{DBLP:journals/jar/Leroy09}. 
Alternatively we can 
provide techniques, supported by tools, to
reason about properties of the bytecode. 
This is what we propose to do in this work.

\paragraph{\bf \itshape Our Contribution.}
We present a complete and formal specification of the EVM in \dafny, available at \url{https://github.com/ConsenSys/evm-dafny}.  
\changes{We provide a formal semantics where the meaning of an instruction 
is given as a partial function
that maps states to states.}
\changes{Our semantics is language-agnostic}, readable and can be used as a 
reference for developers of EVMs  or to aid compiler writers.
\changes{Moreover, it is a complete and usable framework for
formally reasoning about correctness of EVM bytecode using \dafny.}

%% file: background.tex
\section{Background \& Motivation}
\label{sec-background}


In this section we give an overview of the EVM and show how our formal specification in \dafny can be used to verify properties of bytecode programs.

The Ethereum blockchain stores the bytecode of the \emph{contracts} into a database and each contract has its own permanent storage. \changes{In what follows, we assume a given contract and refer to storage as that allocated to this contract}.   
 
\paragraph{\bf \itshape Instructions and States.}   
The EVM~\cite{yp-22} is a \emph{stack-based} machine~\cite{yp-22} which supports 142 instructions:
 arithmetic operations (\eg \lstinline{ADD}, \lstinline{MUL}), comparisons and bitwise operations (\eg \lstinline{ISZERO}, \lstinline{NOT}), 
 cryptographic primitives (\eg \lstinline{SHA3}), environment information (\eg \lstinline{BALANCE}, \lstinline{CALLVALUE}), 
 block information (\eg \lstinline{NUMBER}, \lstinline{GASLIMIT}), stack/memory/control flow (\eg \lstinline{PUSH}, \lstinline{POP}, \lstinline{MSTORE}, \lstinline{SLOAD}, \lstinline{JUMP}), logging (\eg \lstinline{LOG1}), 
 and system operations (\eg \lstinline{CREATE}, \lstinline{CALL}, \lstinline{DELEGATECALL}).  
 An \emph{executing state} of the EVM is a tuple containing several components. 
 We restrict our attention to the following subset of \changes{these} components:
 
\begin{description} 
  \item[code:] a sequence of $n$ bytes indexed
    from $0$ to $n - 1$; The byte at index $0 \leq k < n$ is either an
    instruction {\em opcode} or an {\em immediate operand}.  For instance the
    sequence of opcodes given by $s = [{\tt 0x60, 0x01, 0x60,0x02, 0x01, 0x50, 0x00}]$
    corresponds to the more readable \emph{program}  ``\wlstinline{PUSH1 0x01; PUSH1 0x02;
      ADD; POP; STOP}''.  Here, the byte at $s[1]$ (${\tt 0x01}$)
    is the operand of the instruction \changes{at} $s[0]$
    (\wlstinline{PUSH1}).
  \item[pc:] the \emph{program counter} (initially $0$) identifies the next instruction to execute.  For example, if \wlstinline{pc} is $4$, executing the instruction \changes{at} $s[4]$ (\wlstinline{ADD}) increments it by $1$ so $s[5]$ is the next instruction to execute. When executing instructions with operands (\eg ``\wlstinline{PUSH1 0x01}'' at  $s[0] s[1]$) the \wlstinline{pc} is incremented by $1 + v$ where $v$ is the number of operands. 
  \item[stack:] a \emph{stack} of 256-bit words (initially empty); instructions can push or pop the stack. For example, starting from an empty stack $[\,]$, executing the instructions  ``\wlstinline{PUSH1 0x1; PUSH1 0x2}'' gives \changes{$[{\tt 0x02, 0x01}]$}. Executing the \wlstinline{ADD} instruction from the stack \changes{$[{\tt 0x02, 0x01}]$} pops 2 operands, adds them and pushes the result yielding a new stack $[\tt 0x03 = 0x01 + 0x02]$.
  \item[memory:] a 256-bit addressable, contiguous array of bytes (initially empty).  
  Memory is volatile and only available during the current program execution. Memory expands on-demand when a value is read or written to a given location (which incurs some cost in gas).
  Values can be read from/written to memory using the instructions \wlstinline{MLOAD}, \wlstinline{MSTORE}, \wlstinline{MLOAD8} or \wlstinline{MSTORE8}.
\item[storage:] a map from 256-bit addresses to 256-bit words which constitutes the contract's permanent storage. Storage can be read/written using the instructions \lstinline{SLOAD} or \lstinline{SSTORE}. 
  \item[gas:] the fuel left for future computations. Executing an instruction consumes \emph{gas} in the EVM, and this 
  ensures that no infinite computation can occur.
\end{description} 
In the EVM, program execution may \emph{abort} under exceptional cases including:
\begin{description}
  \item[Out-of-gas:] the gas left in the current state does not cover the cost of executing the next instruction \changes{(including cost of memory expansion if any)};
  \item[Stack exceptions:] the stack size cannot exceed $1024$.  
  Moreover, some instructions (\eg \lstinline{POP}) 
  can only be executed if the stack has enough elements and otherwise the execution should abort.
\end{description} 
The EVM has \emph{failure states} to capture aborted computations. 
As a result, a \emph{state} of the EVM is either a failure state or a non-failure state.

\paragraph{\bf \itshape Bytecode Verification.}

Using our formal semantics, we can guarantee security
properties of bytecode programs using the \dafny verifier. 
\dafny is a verification-friendly language and as such the code can be instrumented with predicates and pre- and postconditions that are checked by the verifier at \emph{compile time}.
We use this feature to prove properties on the bytecode.
%
The following simple \dafny program illustrates a proof:

\begin{lstlisting}
method AddBytes(x: u8, y: u8) {
  //  Initialise an EVM with some gas and the bytecode to execute.
  var st := InitEmpty(gas:=1000, code:=[PUSH1,x,PUSH1,y,ADD]);
  //  Execute 3 compute steps
  st := ExecuteN(st,3);
  //  Check that the top of the stack is the sum of x and y
  assert st.Peek(0) == (x as u256) + (y as u256);
}
\end{lstlisting}

\bigskip 

This simple code snippet illustrates several aspects of the verification process.
First we can verify \emph{family of programs} as the parameters \texttt{x,y} are arbitrary unsigned integers 
over 8 bits.
\changes{This is done by creating an EVM and stepping through the code, \eg using the \lstinline{ExecuteN} function.}
Second, we specify the expected property of the code using the \lstinline{assert} statement (line~7) which is a \emph{verification} statement: it is not {\em executed} at runtime 
as in conventional programming languages but \emph{checked} at \emph{compile-time}, and must hold for all inputs.
\changes{For this program \dafny can prove automatically that the \lstinline{assert} statement is never violated.}
\changes{The proof uses the semantics of opcodes that are invoked in the computation of \lstinline{ExecuteN}.}
Note that if we change \lstinline{u8} to \lstinline{u256} the property does not hold as an overflow can occur in the execution of \lstinline{ADD}: this is flagged by the \dafny verifier with ``Cannot prove assertion at line~7''.
Another set of checks that are performed automatically are related to pre- and postconditions. For instance the 
\lstinline{ADD} instructions requires at least two elements on the stack. 
This is specified by a precondition in the function that defines the semantics of \lstinline{ADD}. 
If the code above had only one \lstinline{PUSH1} instruction \dafny would flag that the \lstinline{ADD} cannot be performed as 
a precondition is violated.
Overall, this short code snippet demonstrates that we can specify and verify functional correctness properties of bytecode, and thanks to the pre- and postconditions used to specify the semantics of the instructions, we can detect/fix possible exceptions (\eg stack overflow) before runtime.

\medskip

The example in listing~\ref{f:inc-proof} shows how we can reason about storage updates and exceptions (aborted computations).



This contract code maintains a counter at storage location $0$ which is incremented by one on every
contract call.  
Initially, the contract storage is unconstrained in the input state \lstinline{st} and,
hence, any location can contain any value.  
The code of the contract aims to capture overflows and 
 to revert if an overflow occurs.
The intent is that either the contract reverts (overflow detected) or the counter is
incremented by $1$.  
Listing~\ref{f:inc-proof} gives a \dafny proof of this.\footnote{The code in the paper may not compile or verify as we have simplified it for clarity. The code in \url{https://github.com/ConsenSys/evm-dafny} compiles and verifies.} 
The preconditions (lines~10--11) ensure that \wlstinline{st}
is an execution (non-failure) state with \wlstinline{pc == 0}, empty stack, enough gas, \changes{and has the contract code to execute.}
 
\medskip 

The postconditions (lines~12-14) specify that the computation either increments the 
counter (at storage location $0$) or the computation reverts.
%
The proof divides up into two essential parts:  
\begin{enumerate}
\item
Execute the first 7 bytecodes and store the intermediate state in  \wlstinline{nst}.
\item 
An overflow occurs when the result of the addition is $0$. 
  So depending on the result at the top of the stack, \lstinline{nst.Peek(0)}, we 
  \changes{decide whether} the rest of the computation will either succeed or revert.
\end{enumerate}
\dafny successfully verifies this code and guarantees the postconditions on lines~12-14 for all input states \lstinline{st} satisfying  the preconditions (lines~10-11).
This provides strong guarantees about the bytecode: 
($i$) it either reverts or computes the increment but never runs out of gas, nor ends up in an invalid state (\eg stack overflow or underflow), 
  ($ii$) the program terminates normally \emph{if and only if} the initial value stored at location $0$ is strictly less than \lstinline{MAX_U256} (line~13), ($iii$) on normal termination, the value in storage location $0$ is incremented by one (line~14).

\begin{lstlisting}[label=f:inc-proof,caption={Verifying bytecode with Reverts.}]
const INC_CONTRACT := Code.Create([
    // Put STORAGE[0] on stack and increment by one
    PUSH1, 0x0, SLOAD, PUSH1, 1, ADD, 
    // If result non-zero branch to JUMPDEST, else REVERT
    DUP1, PUSH1, 0xf, JUMPI, PUSH1, 0x0, PUSH1, 0x0, REVERT,
    // Write result back to STORAGE[0] and return
    JUMPDEST, PUSH1, 0x0, SSTORE, STOP]);

method IncProof(st: State) returns (st': State)
    requires st.OK? && st.PC() == 0 && st.Gas() >= 40000 ...
    requires st.evm.code == INC_CONTRACT
    ensures st'.REVERTS? || st'.RETURNS?
    ensures st'.RETURNS? <==> (st.Load(0) as nat) < MAX_U256
    ensures st'.RETURNS? ==> st'.Load(0) == (st.Load(0) + 1) {
    // Execute upto (and including) JUMPI.
    var nst := ExecuteN(st,7);
    // Consider branches separately
    if nst.Peek(0) == 0 { //  test top of the stack 
        assert nst.PC() == 0xa;
        nst := ExecuteN(nst,3);
        assert nst.REVERTS?;
    } else {
        assert nst.PC() == 0xf;
        nst := ExecuteN(nst,4);
        assert nst.RETURNS?;
    }
    return nst;
}
\end{lstlisting}

%% file: formalisation.tex
\section{The Dafny-EVM}\label{sec-formalisation}

 Our EVM is written in \dafny and provides a definition of the
 semantics as a function mapping states to states.  A key design
 decision made early on was to develop a {\em functionally pure}
 formalisation of the EVM.  In this section we describe the main
 components of the Dafny-EVM and conclude with some observations.

\paragraph{\bf \itshape Machine State.}
Line numbers hereafter refer to Listing~\ref{fig-module-state}.
A state of the EVM is a record containing various fields such as gas,
pc, stack, code, memory.
%
%

\changes{
Each module  (state, stack, memory, \ldots) provides a datatype, possibly incorporating some contraints (\eg \lstinline{EvmState.T}).
}
%
For brevity, we omit some fields which contain
information about the enclosing transaction and the so-called {\em
substate}.  
The \wlstinline{State} datatype (line~7) models normal execution (\wlstinline{OK}), failure (\wlstinline{INVALID}), returning (\wlstinline{RETURNS}), reverting (\wlstinline{REVERTS}), etc.
%


%

%
\begin{figure}[hbtp]
  \centering
  \includegraphics[width=0.825\textwidth]{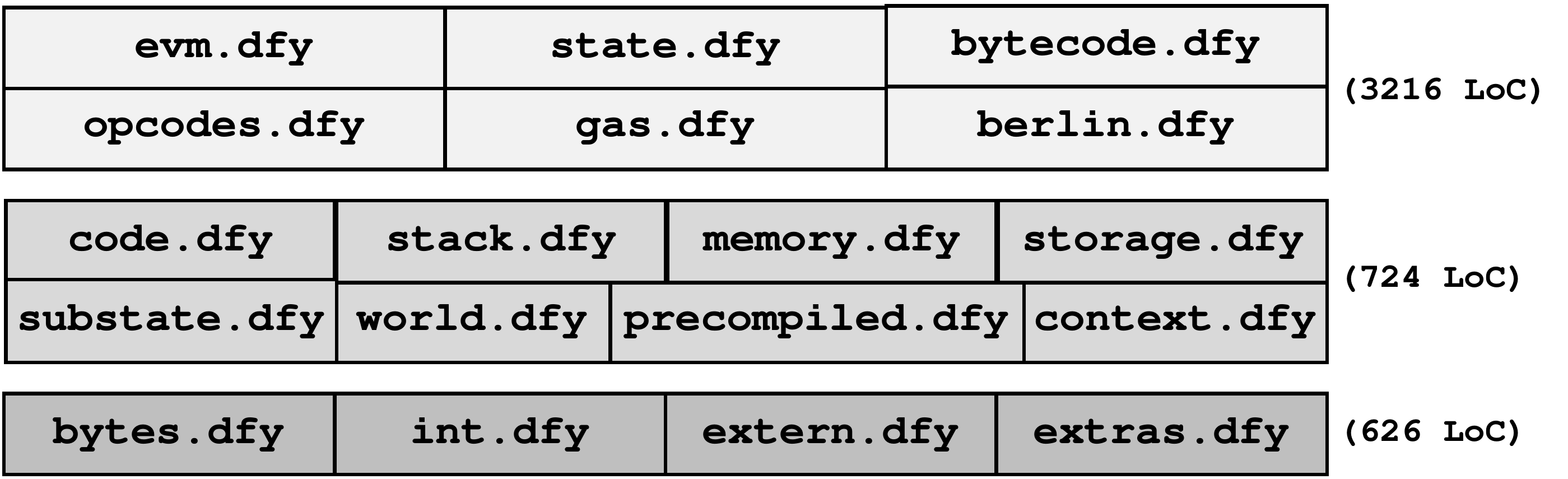}
  \caption{Source files of the Dafny-EVM.  Top group contains bytecode
    semantics and top-level types.  Middle group contains abstractions
    of the main components.  Bottom group are fundamental primitives
    (e.g. for manipulating bytes and ints). ``Loc'' (lines of codes) at the time of writing. }
    \label{PicArc}
  \end{figure}
  

\paragraph{\bf \itshape Stack, Memory and Storage.}  
We have implemented several submodules to provide operations on stack/memory/storage.
This is summarised in \figref{PicArc}. 
We lift the operations on stack/memory/storage into the \lstinline{State} datatype.
In \dafny this is done by adding the functions right after the definition of a datatype (line~11).
This allows us to compose them easily and improves readability.  For instance the \lstinline{Add} function that implements the semantics 
of opcode \lstinline{ADD} is defined using a sequence of operations \lstinline{st.Pop().Pop().Push(...).Next()}  where \lstinline{st} is an executing state (e.g. \wlstinline{OK}).
%
We employ preconditions (\lstinline{requires}) to ensure lifted operations are limited to applicable states only (typically executing states, such as \wlstinline{OK}), and also that preconditions of the functions on stack/memory/storage are satisfied (\eg for \lstinline{Pop()} the stack size must be large enough);
for \lstinline{Push()} (line~20) the stack cannot be full (stack size is limited to 1024).

In \dafny, preconditions 
are checked by the verifier and must provably 
hold at each call site.  
Notice that \dafny enforces the constraints on integer types so every time we compute (\eg \lstinline{ADD}) and store the result in a 256bit word, we must prove that the value is less than $2^{256}$
(the EVM dictates modulo arithmetic for this). 
The pre-/post-conditions and type checks enforced by the \dafny verifier 
help ensure that our EVM specification is consistent and that functions are well-defined.

\begin{lstlisting}[float,floatplacement=!t,caption={Semantics of \lstinline{MLOAD}, \lstinline{Bytecode} module},label=fig-mload]
function method MLoad(st: State) : State
  requires st.IsExecuting() {
if st.Operands() >= 1 then
  var loc := st.Peek(0) as nat;
  var nst := st.Expand(loc,32);         // Break out expanded state
  nst.Pop().Push(nst.Read(loc)).Next()  // Read from expanded state
else
  State.INVALID(STACK_UNDERFLOW)
}
\end{lstlisting}

Memory operations are provided by the \wlstinline{Memory} module, with various functions being attached to
\wlstinline{State}, \eg \lstinline{Read, Write} lines~26--28.
A key observation is that, in both cases, address \wlstinline{addr + 31}
must be within allocated memory.  This is because memory in the EVM is
{\em byte addressable} and we are reading/writing \wlstinline{u256}
values (\ie which are 32 bytes long).  The semantics of
\lstinline{MLOAD} (Listing~\ref{fig-mload}) highlights the complexity
of memory operations.  Since \wlstinline{Read(loc)} (line~6) has the
precondition \lstinline{loc + 31 < Memory.Size} (line~26 of Listing~\ref{fig-module-state}), this must hold for state
\lstinline{nst}.  In fact, this follows because the call to
\wlstinline{Expand()} (line~5) ensures sufficient memory.  If the call
to \wlstinline{Expand()} within \wlstinline{MLoad} was not enforcing
this constraint, then \dafny would raise a precondition violation on
\wlstinline{nst.Read(loc)}.

\paragraph{\bf \itshape Gas.}
In our design, we chose to split out the {\em gas calculation} from
the instruction semantics.  Whilst this does introduce some
repetition, we argue it reduces cognitive load.  In particular, since
this avoids interweaving the gas calculation throughout the
instruction semantics which (for performance reasons) is commonly done
in actual implementations (including the \emph{execution specs}\footnote{\label{execs-specs}\url{https://github.com/ethereum/execution-specs}}).


\paragraph{\bf \itshape Contract Calls.}

Various instructions (e.g. \wlstinline{CALL},
\wlstinline{DELEGATECALL}) enable one contract to call another.  These
differ from others as they can involve executing {\em arbitrarily}
many instructions  in the called contract.  
We implement this
using a mechanism akin to {\em continuations} but, for brevity, omit
the details here.

\begin{lstlisting}[caption=The \lstinline{EvmState} module (partial),label=fig-module-state]
module EvmState {
    datatype Raw = EVM(gas:nat, pc:nat, stack:Stack.T, code:Code.T,
                        mem:Memory.T, world:WorldState.T, ...)
    
    type T = c:Raw | c.context.address in c.world.accounts
    
    datatype State = OK(evm:T) | REVERTS(gas:nat,data:seq<u8>)
        | RETURNS(gas:nat,data:seq<u8>,...) | INVALID(Error) | ...
    {
    //  Predicates 
    predicate method IsExecuting(): bool { ... }

    //  Stack functions 
    function method Capacity(): nat
        requires IsExecuting() { Stack.Capacity(evm.stack) }
    function method Peek(k: nat): u256
        requires IsExecuting() && k < Stack.Size(evm.stack) { ... }
    function method Pop(): State 
        requires IsExecuting() && 0 < Stack.Size(evm.stack) { ... }
    function method Push(v: u256) : State
        requires IsExecuting()
        requires Capacity() > 0 {
            OK(evm.(stack:=Stack.Push(evm.stack,v)))
    }    
    //  Memory functions
    function method Read(address: nat): u256
        requires IsExecuting() && (addr+31) < Memory.Size(evm.mem) {...}
    function method Write(address: nat, val: u256): State
        requires IsExecuting() && (addr+31) < Memory.Size(evm.mem) {...}
    ...
    function method Expand(addr: nat, n: nat): (s': State)
        requires IsExecuting()
        ensures s'.IsExecuting() && MemSize() <= s'.MemSize()
        ensures (addr + n) < MemSize() ==> (evm.mem == s'.evm.mem) {...}
    }
    ... 
}
\end{lstlisting}

\paragraph{\bf \itshape Observations.}\label{sec-observations}

The Dafny-EVM Code provides a readable and executable specification of
the EVM.  There are several benefits of using a verification-friendly
language: using pre- and postconditions to write the semantics
provides a high level of assurance; furthermore, the code is
executable and can be compiled into several target languages including
Java, C\#, Go.  We now highlight some observations based on our
experiences from this project.


\begin{itemize}
\setlength\itemsep{0.5em}  
\item {\bf Specification}.  \dafny treats \wlstinline{function} calls
  within expressions as {\em interpreted}, but treats
  \wlstinline{method} calls as {\em uninterpreted}~\cite{BM07,KS16b}.
  Roughly speaking this means that, when verifying a
  \wlstinline{function} call, the verifier has free access to the
  function's body.  In contrast, for \wlstinline{method} calls,
  the verifier can only
  access what is given in the {\em specification} (i.e. its pre- and
  postconditions).  As such, we consider methods ill-suited for
  formalising specifications (such as for the EVM).  This is because
  we cannot abstract a specification any further than already done
  (i.e. we cannot specify a specification).  
  %

\item {\bf Verification}.  Functions can have preconditions that
  restrict the domain of their inputs.  In \dafny 
  preconditions are
  enforced at each call site. We argue that this results in better
  code by enforcing consistency across function calls.  
  \dafny enforces that every function must have a proof of termination
  which guarantees the absence of infinite loops in our state
  transition function. We believe that this degree of assurance is
  hard to attain with non verification-friendly languages.

\item {\bf Performance}.  Code generated from the functionally pure
  subset of \dafny can perform poorly because of the need to clone
  compound structures (e.g. maps and arrays) to preserve purity
  (\ie referential transparency). 
  \dafny does not, for example,
  employ {\em clone elimination}~\cite{LH11,Shank01,Ode91} or {\em
    mutable value semantics}~\cite{RSZAS22,Pea15c}.
  Performance was not a critical
  concern given our aim of developing a formal
  specification rather than an efficient implementation and
  in practice, we did not encounter any significant issue here.
\end{itemize}
During the project, a number of issues and challenges arose.  For
example, the lack of an exponentiation operator in \dafny meant that,
for the \wlstinline{EXP} bytecode, we had to implement this as a
recursive function.  
\changes{Some low level operations involving bits \& bytes (\eg shifting)
present significant challenges as the native \wlstinline{int} type does not 
support bitwise operators.  }
\changes{One can} use a conversion from 
(e.g. \wlstinline{u256}) into the bitvector types
(e.g. \wlstinline{bv256}) provided by \dafny which do support bitwise
operations \changes{--- however, this can lead to problems verifying code}.

%% file: evaluation.tex
\section{Practical Experiences}

From the outset of this project, we were unsure whether \dafny would be 
\changes{practical}
for this sizeable formalisation task.  Overall, however, we
are pleased to report that \dafny has, for the most part, proven itself
more than capable.  Of course, it was not all plain sailing and we
encountered several challenges which required developing techniques
and/or workarounds.  

\paragraph{\bf \itshape Code Generation.}

\dafny can generate code for a variety of targets, including: C\#, Go,
Java, C++, Python and JavaScript.  
\changes{Furthermore, whilst \dafny does not support I/O operations
  {\em per se}, these can be implemented on the target side.  We took
  advantage of this to embed the \dafny-generated code into a thin Java
  wrapper that performs I/O and allows us to test our EVM against
  existing implementations. Note that the generated code is not
  {\em proved} to be equivalent to the original Dafny code.} %
For various reasons (\eg knowledge
within the team) we chose Java as the target language with \texttt{gradle}
managing the build.  This worked well enough, though there are some
points to make:

\begin{itemize}
\setlength\itemsep{0.3em}    
\item {\bf Foreign Function Interface}.  Code generated from \dafny
  does not conform to the stylistic norms of Java, but is otherwise
  relatively easy to interface with.  A runtime library is provided by
  \dafny against which generated code must be compiled.  
  This provides (amongst other things) alternative
  collection implementations (e.g. \wlstinline{DafnySequence},
  \wlstinline{DafnyMap}, etc).
    
\item {\bf External Code}.  For the semantics of
  \wlstinline{KECCAK256} and some precompiled contracts, we preferred
  to call out to native Java code (i.e. rather than implement
  e.g. \texttt{sha256} in \dafny itself).  However, whilst \dafny does
  support \wlstinline{extern} declarations, these are not (at the time
  of writing) well supported by the Java code generator.  Instead, we
  had to give default implementations (e.g. returning \wlstinline{0})
  and employ build trickery to make it work.  

\item {\bf Target language idiosyncrasies}. Translation to a target language introduces risks. 
E.g., \dafny employs {\em Euclidean Division} for its integer division operator
  (i.e. always rounds {\em down} rather than {\em towards
    zero}), which is a trap for the unwary and by chance
  we identified a bug in the Java code
  generator where sometimes standard division was being
  applied.\footnote{\url{https://github.com/dafny-lang/dafny/issues/2367}}
  We also encountered unsoundness in
  the translation of \dafny collections (e.g. \wlstinline{seq<u8>}) to
  Java\footnote{\url{https://github.com/dafny-lang/dafny/issues/2859}}, and buggy implementation of \wlstinline{datatype} in C\#.
  \footnote{\url{https://github.com/dafny-lang/dafny/issues/1412}}

\end{itemize}

\paragraph{\bf \itshape Verification and Testing.}

For completeness, we developed many unit tests for various components
of our formalism.  The {\em Ethereum Common Tests} also provide \changes{tens of thousands} of tests for ensuring EVM
compatibility.\footnote{\url{https://github.com/ethereum/tests}}
As such, we have been using these to check our formalisation against
existing implementations.  This required generating {\em executable
  code} from our specification which presented several challenges
(discussion of which is unfortunately omitted for brevity).  At the
time of writing, we have selected around \changes{7500 representative tests out of the
13K Common tests (Berlin hardfork)
and 6900 are passing ($92\%$)}. 
Of the 143 failing
tests, the majority (100) are failing because: some precompiled
contracts are not yet fully implemented (44); we do not currently
check for branches into instruction operands (56).
The remaining (approx. 450) tests are skipped for various reasons \eg timeout or breaking the testing system.
\changes{Finally, we note that} all of our tests are run as part of Continuous
Integration before a pull request can be merged.

%% file: relwork.tex
\section{Related Work}\label{sec-related-work}


Initial attempt at a formal specification of the EVM may be attributed to Hirai~\cite{DBLP:conf/fc/Hirai17} with 
a formalisation of the EVM in the programming development environment Lem~\cite{DBLP:conf/icfp/MulliganOGRS14}.
The formalisation in~\cite{DBLP:conf/fc/Hirai17} is restricted to a single contract execution and proving bytecode 
is limited in terms of automation.
Later, Amani~\emph{et al.}~\cite{DBLP:conf/cpp/AmaniBBS18} built upon Hirai's formalisation and proposed an Isabelle/HOL formalisation.
Their contribution introduces a program logic to reason about bytecode (restricted to a subset of 36/142 EVM instructions) but they rely on the construction of a control flow graph to define the semantics of a program.
Reasoning about bytecode is limited to linear sequences of instructions (blocks) and not fully automated.  
Another Isabelle/HOL specification was also developed in~\cite{DBLP:conf/icete/GenetJS20}
specialised for gas consumption analysis and for proving termination of bytecode.

More recently, Grishchenko~\emph{et al.}~\cite{DBLP:conf/post/GrishchenkoMS18} have proposed a partial (not all opcodes are supported and the gas cost semamtics is incomplete) formalisation 
of the EVM in $F^{\star}$ 
targeting \changes{verification of} security properties. 

The most advanced formalisation is probably the KEVM~\cite{DBLP:conf/csfw/HildenbrandtSRZ18}  using the $\mathbb{K}$ Framework~\cite{DBLP:series/natosec/Rosu17}.
It provides a formal and executable specification of the syntax and semantics of EVM bytecode. 
Using the built-in automated tools of the $\mathbb{K}$ Framework, it is possible to generate an interpreter, compiler, debugger and to some extent a verifier that can be used to check the bytecode of some contracts~\cite{deposit-cav-2020}.
The default input format (used for KEVM) of the $\mathbb{K}$ Framework is XML-based which may not be the most developer-friendly 
format. 
%
Similarly, IELE~\cite{iele} attempts to design a more readable language than EVM bytecode and to be the target of high-level languages including Solidity, Vyper, Plutus.  
IELE is defined using the $\mathbb{K}$ Framework and uses LLVM tools (compiler) as a backend.


There are several implementations of the EVM in different languages and clients \eg Geth\footnote{\url{https://geth.ethereum.org}}, Besu\footnote{\url{https://github.com/hyperledger/besu}}, and more recently the \emph{execution-specs} in Python\footref{execs-specs}.
The implementations in Geth and Besu are respectively in Go and Java and cannot be used to reason about bytecode.
The Python implementation relies on specific imperative language features of Python (mutability, exceptions) 
and does not provide a functional definition of the instructions semantics nor an explicit specification of exceptional \changes{cases: for instance the Python code does not provide preconditions or explicit handling of exceptions, and exceptions can happen deep in the call stack which may hinder readability.}


There are several tools 
Oyente~\cite{Oyente}, EtherIR~\cite{EtherIR}, eThor~\cite{DBLP:conf/ccs/SchneidewindGSM20}, Rattle
\cite{rattle-trail-of-bits}, and Certora~\cite{certora-white-paper} to perform static analysis of EVM bytecode.  
There are also extensions to specifically analyse the gas consumption like GASTAP~\cite{Gastap}, GasReducer~\cite{GasReduce}.
Those tools build an abstract representation of the bytecode and it is unclear whether the abstraction is semantics preserving.  

In contrast to the formalisations, implementations and tools referenced above, our formal semantics is
language-agnostic (defines the state transition function as a function), easy to read and developer-friendly,
provides mathematical and verified pre- and postconditions for the semantics of instructions. 
Moroever, our semantics can be used to perform \emph{deductive reasoning} about bytecode 
including gas consumption using standard invariants. 

%% file: conclusion.tex
We have proposed a formal semantics of the EVM in a pure functional subset of \dafny.
Our semantics is {\em human readable}, {\em machine checked} and {\em executable},
and provides a sound framework to formally reason about bytecode.


\noindent This opens up the door for several direct applications:\footnote{Examples are available in \url{https://github.com/ConsenSys/evm-dafny}.}
\begin{itemize}
    \item {complete smart contract verification}: in practice, 
    this can be a cost\-ly process and may require specific verification skills or familiarity with \dafny.
    \item {correctness of compiler optimisations}: several gas optimisation patterns \eg  a sequence of opcodes \lstinline{SWAP1 POP POP} optimised in \lstinline{POP POP} can now be verified.
    \item correctness of under/overflow detection: to detect an overflow in arithmetic modulo \lstinline{ADD(x, y)} it is common to first compute the result \lstinline{r = ADD(x, y)} and then check that \lstinline{r >= x}. 
    We can formally prove that this is sound. 
    \item synthesise verified bytecode: we have designed a methodology~\cite{DBLP:conf/fmics/CassezFQ22} to specify and verify smart contracts directly in \dafny. We are exploring \emph{refinement proof techniques} to synthesise bytecode from the verified \dafny code of a contract. 
    Ultimately we may develop a \dafny-to-EVM \emph{certified compiler}.
\end{itemize}

Although the benefits of our approach are evident in the formal methods' community,
adoption of these techniques in the Ethereum ecosystem is still challenging.
Whilst established techniques, \eg using Solidity to write contracts, or using Python to write specifications, 
can be questionable~\cite{dan-guido-podcast-21}, they are still prevalent in the Ethereum community. 
The main hurdles for mainstream adoption of our approach are probably two-fold: ($i$) provide developer-friendly tools to write contracts; \dafny and the tool support around it (\eg verification performance improvement, counter example generation~\cite{DBLP:conf/tacas/ChakarovFRR22}, VSCode integration) already partially solves this issue; and 
($ii$) educate the Ethereum community to understand the long-term benefits of formal verification for the Ethereum ecosystem.